\documentclass[sigconf, nonacm]{acmart}

\newcommand\vldbdoi{10.14778/3681954.3681992}
\newcommand\vldbpages{3178 - 3191}
\newcommand\vldbvolume{17}
\newcommand\vldbissue{11}
\newcommand\vldbyear{2024}

\newcommand\vldbtitle{\shorttitle} 
\newcommand\vldbavailabilityurl{https://github.com/HKUSTDial/HAIChart}
\newcommand\vldbpagestyle{empty}

\newcommand{\eat}[1]{}

\usepackage{multicol}
\usepackage{latexsym}
\usepackage{amsfonts}
\usepackage{amsmath}
\usepackage{xcolor}
\usepackage{colortbl}
\usepackage{epsfig}
\usepackage{xspace}
\usepackage{graphicx}
\usepackage{subfigure}
\usepackage{paralist}
\usepackage{enumerate}
\usepackage[color,matrix,arrow,all]{xy}
\usepackage{comment}
\usepackage{booktabs}
\usepackage{balance}
\usepackage{stmaryrd}
\usepackage{pifont}
\usepackage{hhline}
\usepackage{listings}
\usepackage{array}
\usepackage{float}
\usepackage[flushleft]{threeparttable}
\usepackage{amsthm}
\usepackage{wrapfig}

\newcolumntype{L}[1]{>{\raggedright\let\newline\\\arraybackslash\hspace{0pt}}m{#1}}
\newcolumntype{C}[1]{>{\centering\let\newline\\\arraybackslash\hspace{0pt}}m{#1}}
\newcolumntype{R}[1]{>{\raggedleft\let\newline\\\arraybackslash\hspace{0pt}}m{#1}}

\usepackage{epsfig}
\usepackage{multirow}
\usepackage{url}

\usepackage{tikz}
\usetikzlibrary{shapes,snakes}
\usetikzlibrary{calc}

\usepackage[lined,boxed,vlined,ruled,linesnumbered]{algorithm2e}

\SetCommentSty{mycommfont}
\usepackage[export]{adjustbox}

\makeatletter
\newcommand{\removelatexerror}{\let\@latex@error\@gobble}
\makeatother

\newcommand{\markt}[1]{\textsf{mark} {#1}}
\newcommand{\enc}[1]{\textsf{encoding} {#1}}
\newcommand{\transf}[1]{\textsf{transformation} {#1}}

\newcommand{\stab}{\vspace{1.2ex}\noindent}
\newcommand{\sstab}{\rule{0pt}{8pt}\\[-2.2ex]}
\newcommand{\vs}{\vspace{1ex}}

\newcommand{\bi}{\begin{itemize}}
	\newcommand{\ei}{\end{itemize}}

\newcommand{\be}{\begin{enumerate}}
	\newcommand{\ee}{\end{enumerate}}
\newcommand{\beqn}{\begin{eqnarray*}}
	\newcommand{\eeqn}{\end{eqnarray*}}

\newcommand{\stitle}[1]{\vspace{1ex}\noindent{\bf #1}}

\newcommand{\etitle}[1]{\vspace{0.8ex}\noindent{\underline{\em #1}}}

\renewcommand{\vs}{{\em vs.}\xspace}
\newcommand{\ie}{{\em i.e.,}\xspace}
\newcommand{\eg}{{\em e.g.,}\xspace}

\newcounter{ccc}

\DeclareMathAlphabet{\pazocal}{OMS}{zplm}{m}{n}

\newcommand{\eop}{\hspace*{\fill}\mbox{$\Box$}\vspace{1ex}}     

\newcounter{example}
\renewcommand{\theexample}{\arabic{example}}
\newenvironment{example}{
\vspace{1ex}
\refstepcounter{example}
{\noindent\bf Example \theexample:}}{
\eop}

\newcommand{\nthesection}{\arabic{section}}

\newcounter{definition}[section]
\renewcommand{\thedefinition}{\nthesection.\arabic{definition}}
\newenvironment{definition}{
\vspace{1ex}
\refstepcounter{definition}
{\noindent\bf Definition {\bf \thedefinition}:}}{\eop
}

\newcounter{alg}[section]
\renewcommand{\thealg}{\nthesection.\arabic{alg}}

\newcounter{arule}
\renewcommand{\thearule}{\arabic{arule}}

\newcounter{claim}
\renewcommand{\theclaim}{\arabic{claim}}

\makeatletter
\newcommand\figcaption{\def\@captype{figure}\caption}
\newcommand\tabcaption{\def\@captype{table}\caption}
\makeatother

\newcommand{\sys}{{\sc HAIChart}\xspace}

\definecolor{shadecolor}{RGB}{200,200,200}

\definecolor{shadecolor1}{RGB}{230,230,230}

\definecolor{shadecolor1}{RGB}{255, 114, 118}

\tikzstyle{mybox} = [draw=black, fill=black!5, thick,
rectangle, rounded corners, inner sep=0pt, inner ysep=6pt]
\tikzstyle{fancytitle} =[fill=black, text=white]

\begin{document}
\title{HAIChart: Human and AI Paired Visualization System}

  
\settopmatter{authorsperrow=4}
\author{Yupeng Xie}
\affiliation{%
	\institution{HKUST (GZ)}
}
\email{yxie740@connect.hkust-gz.edu.cn}

\author{Yuyu Luo*}
\affiliation{%
	\institution{HKUST (GZ) / HKUST}
}
\email{yuyuluo@hkust-gz.edu.cn}

\author{Guoliang Li}
\affiliation{%
	\institution{Tsinghua University}
}
\email{liguoliang@tsinghua.edu.cn}

\author{Nan Tang}
\affiliation{%
	\institution{HKUST (GZ) / HKUST}
}
\email{nantang@hkust-gz.edu.cn}

\begin{abstract}
The growing importance of data visualization in business intelligence and data science emphasizes the need for tools that can efficiently generate meaningful visualizations from large datasets. Existing tools fall into two main categories: human-powered tools (\eg Tableau and PowerBI), which require intensive expert involvement, and AI-powered automated tools (\eg Draco and Table2Charts), which often fall short of {\em guessing} specific user needs. 

In this paper, we aim to achieve the best of both worlds. 
Our key idea is to initially auto-generate a set of high-quality visualizations to minimize manual effort, then refine this process iteratively with user feedback to more closely align with their needs.
To this end, we present \sys, a reinforcement learning-based framework designed to iteratively recommend good visualizations for a given dataset by incorporating user feedback.
Specifically, we propose a Monte Carlo Graph Search-based visualization generation algorithm paired with a composite reward function to efficiently explore the visualization space and automatically generate good visualizations.
We devise a visualization hints mechanism to actively incorporate user feedback, thus progressively refining the visualization generation module. We further prove that the top-$k$ visualization hints selection problem is NP-hard and design an efficient algorithm.
{We conduct both quantitative evaluations and user studies, showing that \sys significantly outperforms state-of-the-art human-powered tools ($\mathbf{21\%}$ better at Recall and $\mathbf{1.8}\times$ faster) and AI-powered automatic tools ($\mathbf{25.1\%}$ and $\mathbf{14.9\%}$ better in terms of Hit@3 and R10@30, respectively).}
\end{abstract}

\maketitle

\pagestyle{\vldbpagestyle}
\begingroup\small\noindent\raggedright\textbf{PVLDB Reference Format:}\\
Yupeng Xie, Yuyu Luo, Guoliang Li, and Nan Tang. \vldbtitle. PVLDB, \vldbvolume(\vldbissue): \vldbpages, \vldbyear.\\
\href{https://doi.org/\vldbdoi}{doi:\vldbdoi}
\endgroup
\begingroup
\renewcommand\thefootnote{}\footnote{\noindent
*Yuyu Luo is the corresponding author. \\
This work is licensed under the Creative Commons BY-NC-ND 4.0 International License. Visit \url{https://creativecommons.org/licenses/by-nc-nd/4.0/} to view a copy of this license. For any use beyond those covered by this license, obtain permission by emailing \href{mailto:info@vldb.org}{info@vldb.org}. Copyright is held by the owner/author(s). Publication rights licensed to the VLDB Endowment. \\
	\raggedright Proceedings of the VLDB Endowment, Vol. \vldbvolume, No. \vldbissue\ %
	ISSN 2150-8097. \\
	\href{https://doi.org/\vldbdoi}{doi:\vldbdoi} \\
}\addtocounter{footnote}{-1}\endgroup

\ifdefempty{\vldbavailabilityurl}{}{
\begingroup\small\noindent\raggedright\textbf{PVLDB Artifact Availability:}\\
The source code, data, and/or other artifacts have been made available at \url{\vldbavailabilityurl}.
\endgroup
}

\section{Introduction}
\label{sec:intro}

\begin{figure}[t!]
	\centering
	\includegraphics[width=\columnwidth]{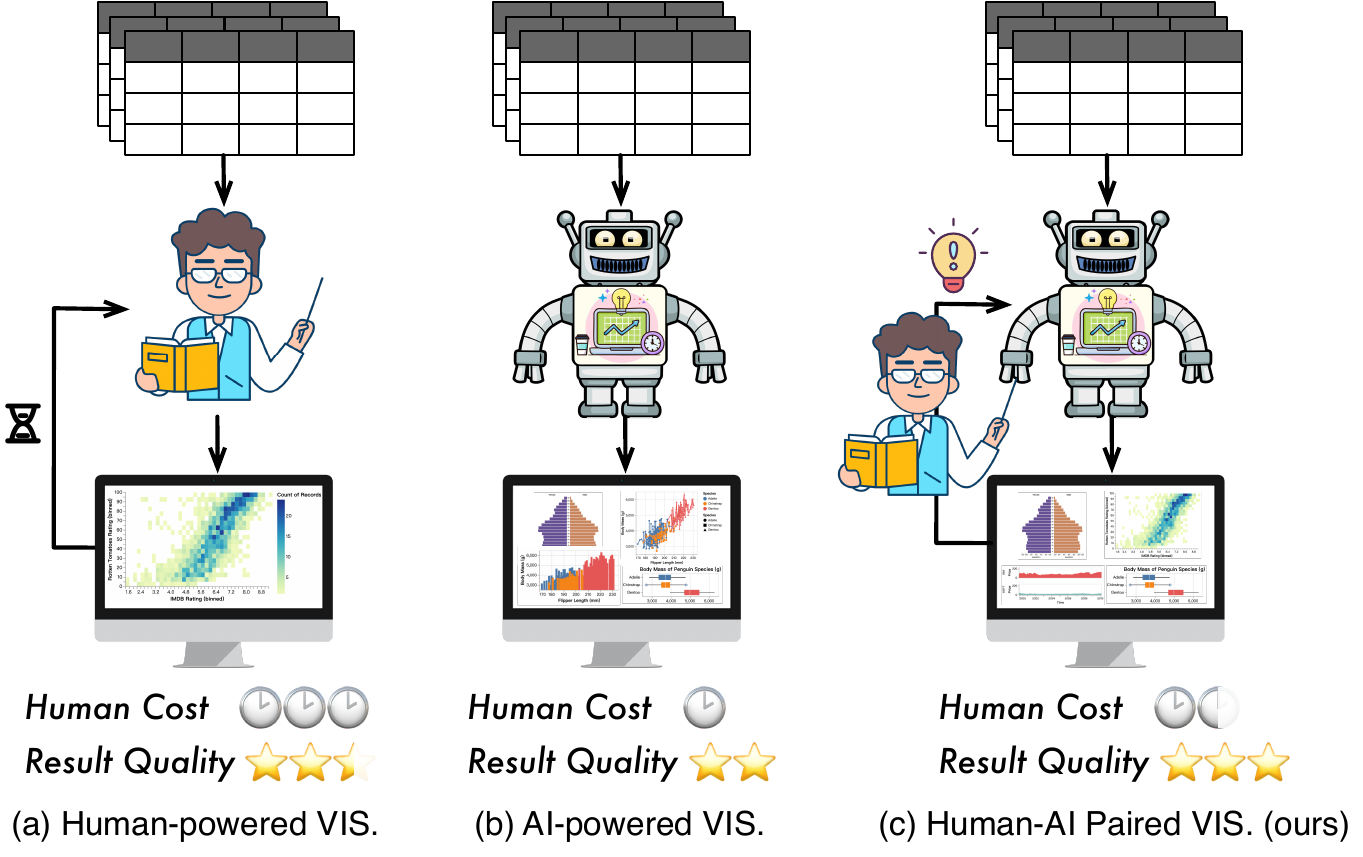}
	\caption{Three cases of visualization approaches}
	\label{fig:example}
\end{figure}

Data visualization is an effective means to uncover underlying insights~\cite{ward2010interactive, DBLP:journals/vldb/QinLTL20, DBLP:journals/tvcg/ShenSLYHZTW23, YE202443}. 
With the increasing importance of data visualization, \textbf{\textit{how to help users effectively and easily create visualizations from massive datasets}} has attracted extensive attention from academia~\cite{wu2021multivision, lee2021lux, DBLP:conf/cikm/ShenST0LW22, DBLP:journals/pvldb/LuoCQ0020, DBLP:conf/icde/LuoCQ0020, DBLP:conf/cidr/0001YF0LH24, DBLP:journals/corr/abs-2406-01265} and industry (\eg Tableau~\cite{tableau}).
Existing visualization tools can be broadly categorized into two groups based on the user effort required for creating visualizations, \ie 
\textit{human-powered} and
\textit{AI-powered}  tools.

\begin{table*}[t!]
    \centering
    \caption{{Comparison of \sys with existing works (Dataset: $D$, User Operations: $U$, Visualizations: $V$)}}
    \label{tab:RelatedeWork}
    \scalebox{0.78}{
        \begin{tabular}{c|c|c|c|c|c|c|c|c}
            \hline 
            \multirow{2}{*}{Types} & \multirow{2}{*}{Systems} & \multirow{2}{*}{Input} & \multirow{2}{*}{Output} & \multirow{2}{*}{Multi-round?} & \multirow{2}{*}{Learned?} & \multicolumn{3}{c}{Recommendation Perspectives} \\ \cline{7-9}
            &  &  &  &  &  & \multicolumn{1}{c|}{Domain Knowl.} & \multicolumn{1}{c|}{User Preferences} & Data Features \\ \hline \hline
            \multirow{3}{*}{Human-powered} & Voyager2~\cite{wongsuphasawat2017voyager} & $D$, $U$& $V$ & \ding{55} & \ding{55} & \ding{51} & \ding{55} & \ding{55} \\ \cline{2-9}
            & DeVIL~\cite{DBLP:conf/cidr/WuPMZR17} & $D$, $U$& $V$ & \ding{55} & \ding{55} & \ding{51} & \ding{51} & \ding{55} \\ \cline{2-9}
            & Lary~\cite{DBLP:journals/cgf/SatyanarayanH14} & $D$, $U$& $V$  & \ding{55} & \ding{55} & \ding{51} & \ding{51} & \ding{55} \\ \hline
            \multirow{10}{*}{AI-powered} & DeepEye~\cite{DBLP:conf/icde/LuoQ0018} & $D$ & $V$ & \ding{55} & \ding{51} & \ding{51} & \ding{55} & \ding{51} \\ \cline{2-9}
            & VizML~\cite{hu2019vizml} & $D$ & $V$ & \ding{55}  & \ding{51} & \ding{55} & \ding{55} & \ding{51} \\ \cline{2-9}
            & Draco-Learn~\cite{2019-draco} & $D$ & $V$ & \ding{55} & \ding{51}  & \ding{51}  & \ding{51}  & \ding{51}  \\ \cline{2-9}
            & Data2Vis~\cite{dibia2019data2vis} & $D$ & $V$ & \ding{55}  & \ding{51}  & \ding{55}  & \ding{55}  & \ding{51}  \\ \cline{2-9}
            & KG4VIS~\cite{li2021kg4vis} &$D$, Knowl. Graph
& $V$ & \ding{55} 
            & \ding{51} & \ding{51}  & \ding{55}  & \ding{55}  \\ \cline{2-9}
            & Table2Charts~\cite{zhou2021table2charts} & $D$ & $V$ & \ding{55}  & \ding{51} & \ding{55} & \ding{55} & \ding{51} \\ \cline{2-9}
            & VizGRank~\cite{DBLP:conf/dasfaa/GaoHJZW21} & $D$ & $V$ & \ding{55}  & \ding{55} & \ding{51} & \ding{55} & \ding{51} \\ \cline{2-9}
            & PVisRec~\cite{qian2022personalized} & $D$ & $V$ & \ding{55}  & \ding{51} & \ding{55} & \ding{51} & \ding{51} \\ \cline{2-9}
            & {LLM4Vis~\cite{DBLP:conf/emnlp/WangZWLW23}} & {$D$} & {$V$} & {\ding{55}} & {\ding{51}} & {\ding{55}} & {\ding{55}} & {\ding{51}} \\
            \cline{2-9}
            & {PI2~\cite{DBLP:conf/sigmod/Chen022}} & {$D$, SQL Queries} & {$V$, Interface} & {\ding{51}} & {\ding{55}} & {\ding{51}} & {\ding{51}} & {\ding{55}} \\
            \hline\hline
            \begin{tabular}[c]{@{}l@{}}Human-AI Paired\end{tabular} & \sys  & $D$, User Feedback & $V$, Hints& \ding{51} & \ding{51} & \ding{51} & \ding{51} & \ding{51} \\ \hline
        \end{tabular}
    }
\end{table*}

\stitle{Human-powered Visualization} tools, including Voyager2~\cite{wongsuphasawat2017voyager}, DeVIL~\cite{DBLP:conf/cidr/WuPMZR17}, and Lary~\cite{DBLP:journals/cgf/SatyanarayanH14}, enable users to create their desired visualizations by manually specifying the data attributes, transformations, and visual encodings, as shown in Figure~\ref{fig:example}(a). 
This process is time-consuming and error-prone because it requires a deep understanding of the datasets and analytical tasks. Users often need to attempt multiple iterations to achieve the final visualization. Consequently, interactive visualization demands \textit{domain-} and \textit{data-specific} expertise, which can significantly hinder novice users from effectively engaging in visual analysis. 

In response to the above challenges, recent research tries to \textit{automate} the visualization process empowered by artificial intelligence.

\stitle{AI-powered Automatic Visualization} aims to enumerate and recommend the (top-$k$) best visualizations for a given dataset based on predefined constraints~\cite{DBLP:conf/icde/LuoQ0018, li2021kg4vis, deng2022dashbot, ncnet, 2019-draco} or learning-based recommendation algorithms~\cite{qian2022personalized, zhou2021table2charts}.
For example, DeepEye~\cite{DBLP:conf/icde/LuoQ0018} recommends visualizations based on visualization rules and a ranking model, while Table2Charts~\cite{zhou2021table2charts} generates top-$k$ visualizations using a table-to-sequence generation model. Although these tools provide a valuable starting point for data analysis by suggesting ``good'' visualizations, they risk misleading users with indiscriminate recommendations. The key issue is that these tools typically rely on static logical rules or deep learning models without adequately capturing user intent or feedback, hindering the tailoring of recommended visualizations to better suit users' needs.

\stitle{Human and AI Paired Visualization.} 
%
To mitigate the limitations of the above methods and strike a balance between human-powered and AI-powered visualization, we present \textit{Human and AI Paired Visualization}, as shown in Figure~\ref{fig:example}(c). Our key idea is to \textit{initially} recommend a set of \textit{high-quality} visualizations and then continue to refine these recommended visualizations by providing users with \textit{visualization hints} for further tuning.
This approach not only simplifies the process for novice users seeking to discover effective visualizations but also empowers users to actively participate in and guide the AI-powered visualization process through the human-centered feedback, as shown in Table~\ref{tab:RelatedeWork}.

Note that another natural way humans interact with AI for visualizations is through conversations powered by Large Language Models (LLMs), or briefly LLM4VIS~\cite{cheng2023gpt, DBLP:conf/acl/Dibia23, tian2023chartgpt, DBLP:conf/emnlp/WangZWLW23,he2024leveraging, kim2023good}.
However, the key limitation of LLM4VIS is that even with user natural language hints,
it is hard to precisely modify visualizations outputted by LLMs using natural language; that is, easy-to-use but hard-to-calibrate. Please refer to Section~\ref{sec:experiment} for an empirical comparison with LLM4VIS powered by GPT-4 and Section~\ref{sec:related} for more discussions.

\stitle{Challenges.} There are three challenges in our problem.
{\bf (C1)} How can we effectively and efficiently explore the visualization search space to recommend visualizations that are both high-quality and relevant?
{\bf (C2)} How can we evaluate the ``{\em goodness}'' of a generated visualization comprehensively?
{\bf (C3)} How can we understand and effectively integrate user feedback to guide the system toward visualizations that align with users' requirements?

\stitle{Our Proposal: \sys.}
To address these challenges, we propose \sys to automatically and progressively recommend a series of visualizations that adapt to evolving user preferences over time. This process is characterized by the system's ability to learn from user interactions and feedback, iteratively refining the recommended visualizations. \textit{The objective is to converge towards a final set of visualizations that closely align with the user's preferences, within the constraints of the available data and visualization search space.} This iterative optimization is essentially a series of decisions based on user intent, similar to the key idea in reinforcement learning. 

We adopt the reinforcement learning framework to implement \sys.
We treat a visualization as a sequence of operations, such as chart types and $x$/$y$-axes configurations. 
Thus, the ``state'' is the current sequence of \textit{incomplete} visualization, the ``action'' refers to selecting the subsequent visualization operation within that state, and the ``environment'' is the visualization system itself. The environment evaluates the quality of the visualizations and provides rewards to guide the ``agent'' responsible for generating and recommending visualizations.

We utilize the Monte Carlo Graph Search (MCGS) algorithm to build this agent, enabling it to efficiently navigate the extensive search space to discover optimal visualizations (addressing {\bf C1}).
To more accurately evaluate the visualizations, we design a composite reward function that takes into account the data features, visualization domain knowledge, and user preferences to ensure the quality of the visualizations (addressing {\bf C2}). 
Moreover, we devise a visualization hints selection algorithm for recommending useful hints (\eg ``{\sf explore why flights are delayed}'') to assist the user in data exploration. This approach can integrate user feedback for better visualization results (addressing {\bf C3}).




\begin{figure*}[t!]
	\centering
	\includegraphics[width=\textwidth]{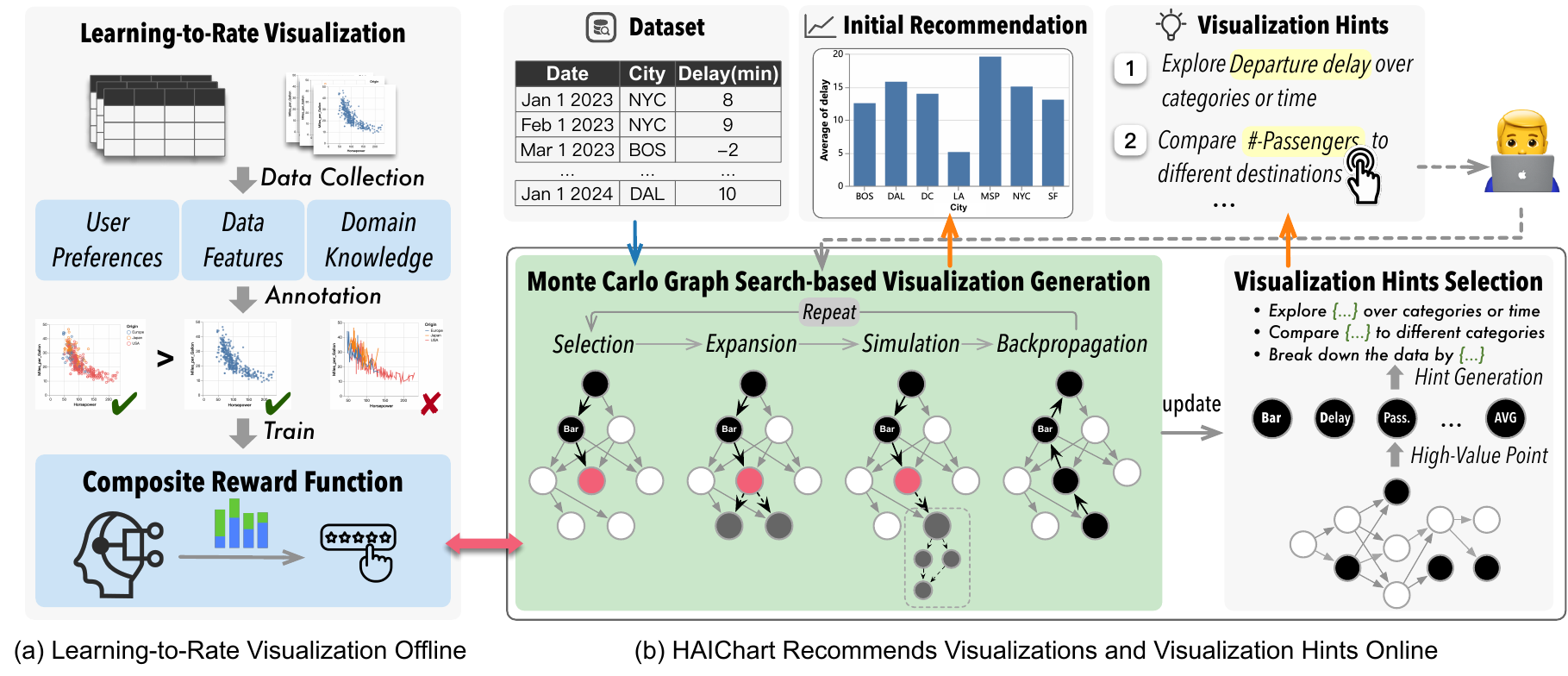}
 \vspace{-2em}
	\caption{An overview of \sys}
	\label{fig:overview}
\end{figure*}

\stitle{Contributions.} 
We make the following notable contributions.

\sstab (1) {\sf Problem Statement}. We formally define the problem of human and AI paired visualization (Section~\ref{sec:preliminary}).

\sstab (2) {\sf \sys}.
We propose a reinforcement learning-based system, \sys, to integrate human insights and AI capabilities for better visualizations. \sys initially generates a set of promising visualizations for users and iteratively improves the alignment of recommended visualizations with user preferences by selecting a set of visualization hints for interaction (Sections~\ref{sec:overview} \& \ref{sec:rl}).
 
\sstab (3) {\sf Learning-to-Rate Visualization}. We design a composite reward function to rate the quality of visualizations, which is used to accurately guide the visualization generation process (Section~\ref{sec:reward}).

\sstab (4) {\sf Visualization Hints Selection}.  We design visualization hints as the proxy to collect user feedback, prove the selection of the top-$k$ visualization hints is an NP-hard problem and design an efficient algorithm to select the top-$k$ hints (Section~\ref{sec:recommendation}).

\sstab (5) {\sf Experiments}. 
{We conduct both quantitative evaluations and user studies to demonstrate that \sys outperforms human-powered tools in terms of both accuracy {($21\%$ better at Recall)} and efficiency in creating visualizations ($1.8\times$ faster) and is better than AI-powered systems in effectiveness by {25.1\% and 14.9\% in terms of Hit@3 and R10@30, respectively} (Section~\ref{sec:experiment}).}

\section{Problem and Solution Overview}
\label{sec:preliminary}

\subsection{Problem}
Given a relational table $D$, let $\mathcal{V}$ be all valid visualizations, and $\mathbf{V}^+$ be a set of ``good'' visualizations, both being associated with $D$. In practice, this set of visualizations $\mathbf{V}^+$ is often manually selected by the users, such as the visualizations in a dashboard.

\stitle{Visualization Selection.}
The problem of {\em visualization selection} is to select a set $\mathbf{V}$ of visualizations from all valid visualizations $\mathcal{V}$, which maximizes $\mathbf{V} \cap \mathbf{V}^+$ (\ie maximizes the number of good visualizations being selected) and minimizes $\mathbf{V} \setminus \mathbf{V}^+$ (\ie minimizes the number of bad visualizations being selected).

When being operated manually, visualization selection is time-consuming, and its effectiveness depends on the user's expertise in data visualization and domain knowledge of specific datasets.


To reduce human effort, visualization recommendation algorithms have been studied. These algorithms are mostly powered by machine learning algorithms. However, these recommendation algorithms are mainly one round.

\stitle{One-Round Visualization Recommendation.}
The problem of {\em one-round visualization recommendation} is to automatically suggest a set $\mathbf{V}^S$ of visualizations and ask the user to select a subset $\mathbf{V} \subseteq \mathbf{V}^S$. 

In practice, a one-round visualization recommendation may not be sufficient to support the user's diversified intent, \eg selecting all charts for a dashboard.
Therefore, we propose to study a new problem, which does multiple rounds of interactions with users to collectively select all required good visualizations.

\stitle{Multi-Round Visualization Recommendation.}
The problem of {\em multi-round visualization recommendation} is to recommend $n$ rounds of visualizations to users. In each round, we can suggest $\mathbf{V}_i^S$ visualizations (optionally with some natural language hints), and the user can select a subset $\mathbf{V}_i \subseteq \mathbf{V}_i^S$ visualizations in the $i$-th round. After $n$ rounds, the user will select $\mathbf{V} = \bigcup_{i=1}^n \mathbf{V}_i$ visualizations.


\subsection{Solution Overview} 
\label{sub:overview}

\stitle{HAIChart Overview.} 
As shown in Figure~\ref{fig:overview}, \sys consists of two components: an offline part for learning-to-rate visualizations and an online part for automatically recommending good visualizations and providing visualization hints for user interaction.

\stitle{{Offline: Learning-to-Rate Visualization.}}
As shown in Figure~\ref{fig:overview}(a), the offline part is responsible for understanding and evaluating visualization quality. To achieve this, we incorporate visualization rule of thumb, data features, and user preferences to learn a composite reward function.

\stitle{Online: Multi-Round Visualizations and Hints Recommendation.}
As shown in Figure~\ref{fig:overview}(b), \sys leverages an \textit{agent} powered by Monte Carlo Graph Search to recommend a set of promising visualizations along with visualization hints to guide further exploratory actions.
Specifically, this agent initially traverses the visualization search space effectively and efficiently. It generates possible visualizations, estimates their quality based on the well-trained composite reward function, and finally returns a ranked list of high-quality visualizations.
In addition, the system computes visualization hints, derived from the current states of the Monte Carlo Graph. The user can browse the recommended visualizations and hints. Once the user selects a hint, the system will further explore the visualization space and recommend visualizations guided by the selected hint.

\section{Reinforcement Learning for Multi-\mbox{Round Visualization Recommendation}}
\label{sec:overview}
 
\subsection{Preliminary}
\label{sub:pre}

Inspired by previous works~\cite{2019-draco, DBLP:conf/icde/LuoQ0018, ncnet}, we adopt a visualization query to represent all possible visualizations discussed in this paper.

\stitle{Visualization Query.} 
Given a relational table $D$ with the scheme
$R(A_1,...,A_m)$, a visualization query $Q$ is composed of a sequence of visualization operations (\ie \textit{actions})  $\mathcal{A} (\markt, ~\enc, ~\transf)$. Specifically:
\bi
\item \markt represents the type of visualization charts including bar, line, pie, and scatter.
\item \enc denotes the encoding of data attributes including $x$/$y$-axes, aggregate functions, and color schemes.
\item \transf defines data transformations: filtering, binning, grouping, sorting, and  top-$k$ operations.
\ei

Applying the visualization query $Q$ to the table $D$ results in the corresponding visualization $V$, \ie $V = Q(D)$.
Therefore, any visualization $V$ for a table $D$ can be represented by a combination of data attributes and cell values in $D$, along with visualization operations (\ie actions) in $\mathcal{A}$. 

\begin{figure}[t!]
	\centering
	\includegraphics[width=1.0\columnwidth]{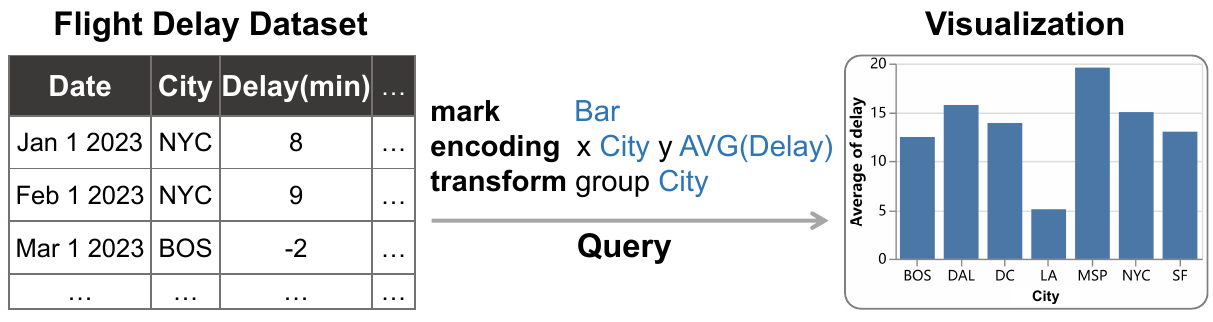}
	\caption{A visualization query}
	\label{fig:visualization_query_results}
\end{figure}

\begin{figure}[t!]
	\centering
	\includegraphics[width=0.9\columnwidth]{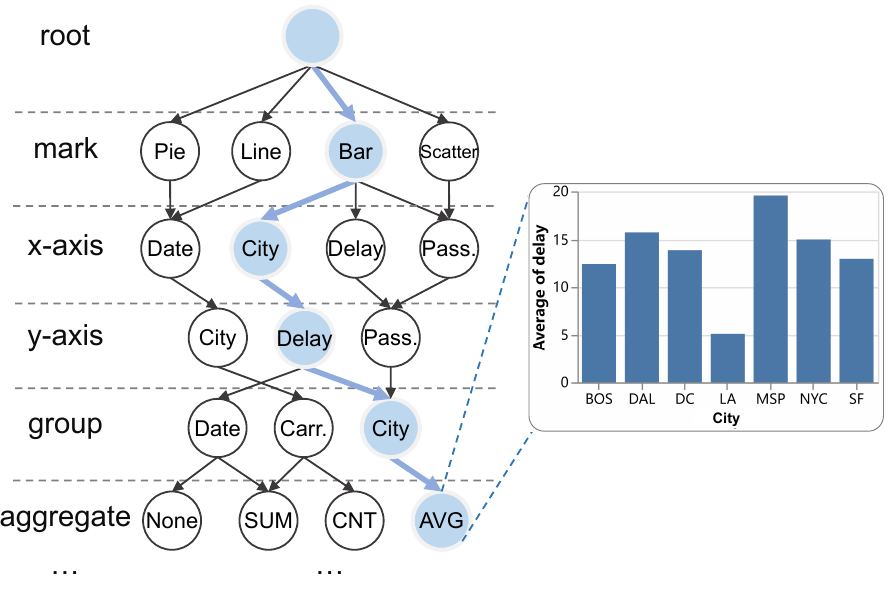}
	\caption{A visualization query graph}
	\label{fig:visualization query graph}
\end{figure}

\begin{example} [Visualization Query]
    Figure~\ref{fig:visualization_query_results} shows an illustrative query ($Q$) applied to the flight delays dataset ($D$).
    The query visualizes a bar chart with the $x$-axis encoding cities (\ie~ {\sf City} column) and the $y$-axis showing the average delay (\ie ~{\sf AVG(Delay)}). 
    The bar chart ($Q(D)$) shows the trend in average delays, showing that LA (\ie Los Angeles) has the shortest average delays, while MSP (\ie Minneapolis) has the longest average delays.
\end{example}

Undoubtedly, the search space for visualizations is very large, growing exponentially with the increase in number of columns and combinations of visualization operations.

To effectively navigate this huge search space, we introduce the concept of \textit{visualization query graph} to represent all possible visualizations for a given dataset, defined as follows:


\begin{definition} [Visualization Query Graph]
    Given a table $D$, the visualization query graph $\mathcal{G(V, E)}$ is defined as a directed acyclic graph. Specifically, each node $v \in \mathcal{V}$ represents a visualization operation, and each directed edge $e \in \mathcal{E}$ indicates a transition from node $v_i$ to $v_j$. 
   The weight of an edge $(v_i, v_j)$, denoted by $w_{ij}$, indicates the effectiveness of transitioning from operation $v_i$ to $v_j$.
  A path from the root node to an end node within this graph represents a sequence of visualization operations, which together form a candidate visualization query.
\end{definition}

\begin{example} [Visualization Query Graph]
    As illustrated in Figure~\ref{fig:visualization query graph}, each layer of the graph corresponds to the possible values that a visualization action can take. For example, the chart type (\textit{mark}) could be \textit{bar}, \textit{line}, \textit{scatter}, \textit{pie}, {\em etc}. Taking the blue path as an example, it represents a visualization query -- ``{\bf mark} Bar {\bf encoding} {\sf x} City {\sf y} AVG(Delay) {\bf transform} {\sf group} City''. After executing this query over the dataset, it will create a bar chart.   
%
\end{example}

\begin{figure*}[t!]
	\centering
	\includegraphics[width=1\textwidth]{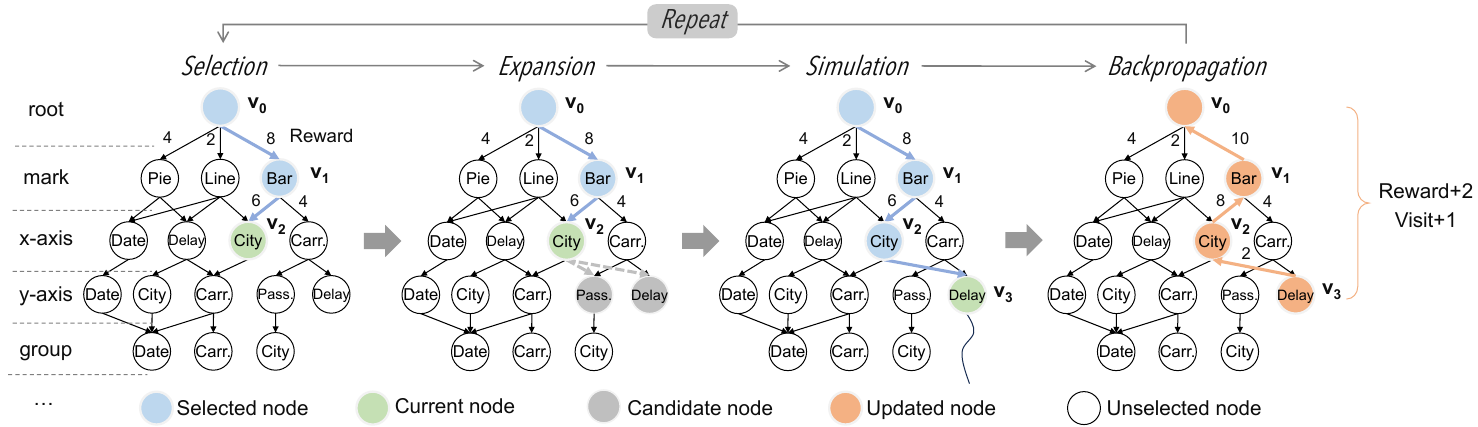}
	\caption{An example of MCGS-based visualization generation}
	\label{fig:MCGS-based_Visualization_Generation}
\end{figure*}

\begin{figure}[t!]
	\removelatexerror
	{\small
		\begin{algorithm}[H]
    \caption{MCGS-based Visualization Generation \label{alg:mcgs_process}}
    \KwIn{Dataset $D$;}
    \KwOut{Top-$k$ Visualization Results $R = \{V_1, ..., V_k\}$;}
    $node \leftarrow$ Initialize($D$); 
    $\mathcal{G} \leftarrow \{node\}$; 
    $R \leftarrow \{\}$;

    \For{each iteration}{
        Set a visualization query $Q = \text{``''}$; \\
        $node = root$;

        \While{$Q$ is invalid}{
            \uIf{node is fully expanded}{
                $node \leftarrow$ select($\mathcal{G}$, $node$);\\
                Add $node$ to $Q$;\\
                \If{$Q$ is valid}{
                    break;
                }
            }
            \Else{
                $node \leftarrow$ expand($\mathcal{G}$, $node$);\\
                
                $Q \leftarrow$ simulation($\mathcal{G}$, $node$);\\
                }
            }
        $score \leftarrow$ reward($Q$);\\
        BackPropagation($score$, $Q$, $\mathcal{G}$); \\
        $V \leftarrow$ $Q(D)$; // Corresponding visualization \\ 
        add $V$ to $R$;  
    }
    \Return $R$;
\end{algorithm}
}
\end{figure}

\subsection{\sys Details} 
\label{sub:train} 

Given a dataset for visualization, users typically select visualization types, data columns, and transformation operations based on their preliminary understanding. If the results do not meet their analysis requirements, they adjust their operations to optimize the output. This visualization process is a sequence of decision-making based on user intent, similar to Reinforcement Learning (RL), where feedback from the visualization system (\ie environment) helps optimize strategies to achieve satisfactory outcomes.

Therefore, we model the problem of visualization generation and recommendation as a Markov Decision Process (MDP)~\cite{bellman1957markovian} and implement it using an RL framework.

The key RL components of \sys are as follows: 

\stitle{State.} 
To apply the RL framework to our problem, it is crucial to accurately define the state, which serves as input to the agent and aids in making decisions during visualization generation. We transform the visualization generation problem into a process that starts from an initial state and proceeds through a series of visualization operation/action decisions to reach a target state. Each state corresponds to a visualization query consisting of multiple visualization clauses, as described in Section~\ref{sub:pre}.

Specifically, visualization queries can be classified into two types:

\stab(1) {\em Partial Visualization Queries}.
These queries need to be further extended to form a complete query and reflect historical decision paths in Visualization Query Graph. For example, an incomplete query might have specified the type of chart (\eg bar chart) but not yet selected a specific data field for the $x$- or $y$-axis.

\stab(2) {\em Complete Visualization Queries}.
These queries do not require additional extensions and can be evaluated for visualization quality by our reward function. After that, the system updates the nodes and edges of the graph based on the evaluation results.

\stitle{Action.} An action is an operation that an agent can perform based on its current state, determining the next step in a visualization query sequence. As discussed in Section~\ref{sub:pre}, the visualization query can be categorized into three main types of actions: {\tt mark}, {\tt encoding}, and {\tt transformation}. Therefore, for a given dataset $D$, the available action space $\mathcal{A}$ is fixed.


\stitle{Agent.} The agent combines the Monte Carlo Graph Search (MCGS) algorithm and the Upper Confidence Bound (UCB)~\cite{auer2002finite} algorithm to generate visualizations and make decisions. The MCGS performs well in navigating complex and uncertain search spaces, while the UCB algorithm effectively balances the trade-off between exploiting known optimal solutions and exploring new possibilities, thereby optimizing the agent's decision-making process (see Section ~\ref{sec:rl}).

\stitle{Reward.} The reward function evaluates the quality of the visualizations created by the MCGS algorithm and updates the visualization query graph accordingly. This guides the agent to generate higher quality visualizations. The reward function considers three key factors: data features, visualization domain knowledge, and user preferences to accurately assess visualization quality (see Section~\ref{sec:reward}).


\stitle{Environment.} The environment in \sys first uses a pruning algorithm to efficiently generate valid visualization queries during the generation process. Second, it evaluates the quality of valid queries and returns a reward to guide the training process.



\section{Visualization Generation}
\label{sec:rl}



\subsection{Monte Carlo Graph Search-based Visualization Generation}
\label{sec:mcgs_vis_gen}
In Section~\ref{sub:pre}, we introduced the visualization query graph, where each visualization query is represented by a path within the graph. All possible paths within the query graph constitute the search space for visualizations. We can use search algorithms to explore the graph and find high-quality visualizations. However, as the dataset size increases, the search space expands exponentially, making it challenging to efficiently find good visualization results.

Traditional graph search methods like Depth-First Search (DFS) and Breadth-First Search (BFS) perform well in structured search spaces but often fall short in large and complex spaces because they cannot dynamically adjust the search strategy based on the information already found. To address these limitations, we propose a Monte Carlo Graph Search-based algorithm to navigate the vast search space efficiently.


\stitle{Key Idea.}
Our key idea is to dynamically accumulate and utilize shared information from nodes throughout the search process. Initially, shared information is limited, but as the search progresses, the accumulated information from explored nodes becomes increasingly significant, providing greater support for decision-making. This allows the algorithm to gradually reduce its reliance on random simulations, leveraging accumulated knowledge to guide the search and enhance efficiency~\cite{DBLP:conf/acml/LeurentM20}.

We first overview our Monte Carlo Graph Search-based visualization generation algorithm. As shown in Algorithm~\ref{alg:mcgs_process}, the algorithm begins by taking a dataset $D$ as input, aiming to generate the top-$k$ visualization results $R$. In the initial phase, the algorithm initializes a root node based on the dataset $D$, which serves as the starting node of the graph $\mathcal{G}$ (Line 1).

In each iteration, the algorithm starts by initializing a visualization query $Q$ (Line 3) and then starts exploring from the root node. During the node selection phase, each choice is added to the visualization query $Q$ (Line 7-8) until an unexpanded node is encountered, followed by an expansion of this node (Line 12). When the visualization query $Q$ is valid, the algorithm utilizes a reward function to calculate the score of the query (Line 14), then performs backpropagation to update the information on the graph $\mathcal{G}$ (Line~15). Finally, the query is transformed into the corresponding visualization results and stored in $R$ (Lines 16-17). 
After $k$ visualizations are selected, the algorithm returns this set $R$.


Next, we will detail the procedure of Algorithm~\ref{alg:mcgs_process} through Figure~\ref{fig:MCGS-based_Visualization_Generation}. The algorithm mainly includes the following four steps: 

\stitle{Selection.} In the selection phase (Line 7 in Algo.~\ref{alg:mcgs_process}), the algorithm starts from the root node and recursively selects the optimal child nodes until it reaches a node that has not yet been fully expanded. As shown in Figure~\ref{fig:MCGS-based_Visualization_Generation}, the blue nodes represent the nodes selected in this phase, while the green nodes represent the currently selected nodes that have not been fully expanded. To effectively utilize the feedback information provided by the reward function and balance exploration and exploitation, we adopt the Upper Confidence Bound (UCB) algorithm. The key idea of this algorithm is to select the child node with the highest confidence upper bound at each iteration. The specific selection strategy of the UCB algorithm can be represented by the following formula:

\begin{equation}
    UCB = \begin{matrix}
    \underbrace{\bar{X}_i} \\ \textit{exploitation} 
   \end{matrix}
 +  \begin{matrix}
 	\underbrace{c \sqrt{{2 \ln n}/{n_i}}} \\ \textit{exploration} 
 \end{matrix}
\end{equation}%
where $ \bar{X}_i $ is the average reward of child node $ i $, $ n $ is the number of visits to the current node, $ n_i $ is the number of visits to child node $ i $, and $ c $ is a constant used to balance exploration and exploitation.


\stitle{Expansion.} During the expansion phase (Line 12 in Algo.~\ref{alg:mcgs_process}), the algorithm selects the next valid action based on the current state (\ie an incomplete visualization query). Specifically, it removes those low-quality visualizations that are either syntactically incorrect or violate visualization rules (see Section~\ref{sec:optim}). 
For example, when processing node $v_2$ (in Figure~\ref{fig:MCGS-based_Visualization_Generation}), the algorithm identifies two high-benefit candidates and randomly selects one for further expansion.

\stitle{Simulation.} During the simulation phase (Line 13 in Algo.~\ref{alg:mcgs_process}), the algorithm starts at the current node (\eg node $v_3$ shown in Figure~\ref{fig:MCGS-based_Visualization_Generation}) and performs simulation actions. 
First, it checks whether the current visualization query is valid. If the query is invalid, the algorithm randomly explores the next node according to the visualization grammar rule until a valid query is constructed. Once a valid query is formed, the learning-to-rate visualization part assigns a reward.

\stitle{Backpropagation.} This phase (Line 15 in Algo.~\ref{alg:mcgs_process}) propagates the simulation results through the visualization query graph. 
After the simulation phase, as shown in Figure~\ref{fig:MCGS-based_Visualization_Generation}, the score obtained by node $v_3$ is propagated along its path.
%
Each node and edge along the path updates its reward value and visit count based on the simulation outcomes. Our learning-to-rate visualization module evaluates the visualization quality and updates the node information along the search path, guiding the algorithm to make more precise decisions in subsequent explorations by feeding the results of each simulation back into the visualization query graph.

\etitle{Terminal Condition.} The above steps are repeated until the maximum number of iterations is reached.

\begin{figure}[t!]
	\centering
	\includegraphics[width=1.0\columnwidth]{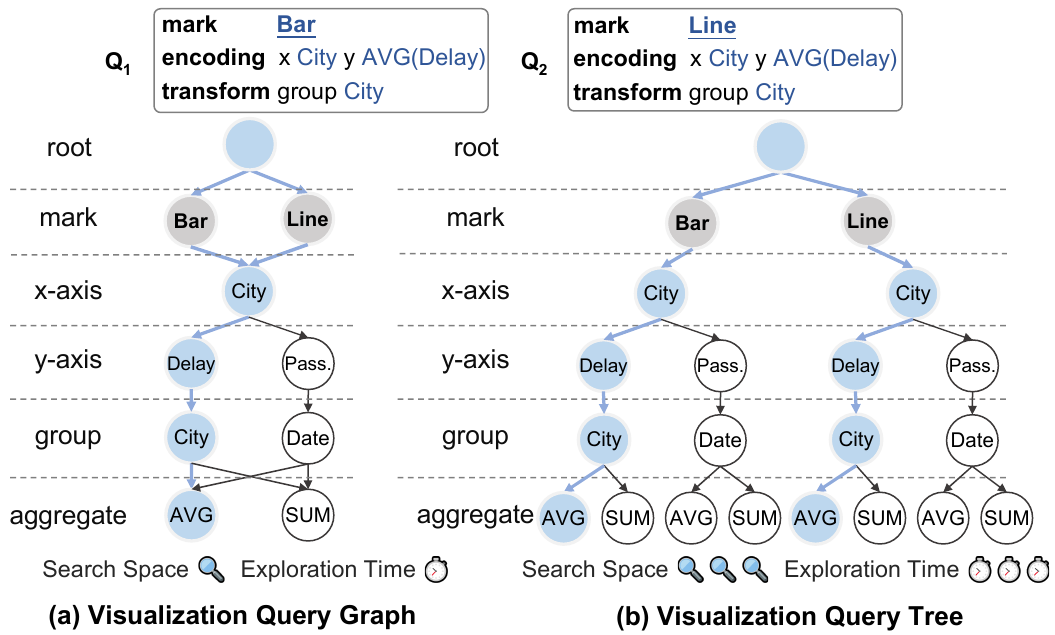}
	\caption{Comparison of tree and graph structures}
	\label{fig:tree_graph}
\end{figure}

\stitle{{Comparison of MCGS and MCTS.}}
{The Monte Carlo method solves complex problems through random sampling but requires simulation steps for each new node, impacting performance based on node count~\cite{robert1999monte}. 
Traditional MCTS methods rely on extensive, random, and time-consuming simulations to determine the next step~\cite{DBLP:journals/tvcg/ShiXSSC21}. Unlike traditional MCTS, our MCGS algorithm explores a graph structure instead of a tree structure, allowing for more effective information sharing between nodes and reducing the number of nodes, thereby increasing search efficiency~\cite{DBLP:conf/cig/TotCLD22}.}


As shown in Figure~\ref{fig:tree_graph}, visualization queries $Q_1$ and $Q_2$ differ only in chart type; other clauses are identical. In a tree structure (Figure~\ref{fig:tree_graph}(b)), this information cannot be shared due to the strict parent-child hierarchy, leading to redundancy. In contrast, the graph structure (Figure~\ref{fig:tree_graph}(a)) allows node information sharing across different queries, reducing redundancy and improving search efficiency.

More concretely, we compare the number of nodes between MCGS and MCTS. Consider 4 chart types, a dataset with $m$ columns, encoding that considers both the 
$x$- and $y$-axes, and data transformations that include grouping and 4 aggregate operations. In MCTS, each path is independent, resulting in a total node count of \( N_{\text{tree}} = 4 \times m^3 \times 4 = 16m^3 \). In MCGS, using the graph structure to share nodes, the total node count is \( N_{\text{graph}} = 4 + 3 \times m + 4 = 3m + 8 \). The reduction factor is  \(R = \frac{N_{\text{tree}}}{N_{\text{graph}}} = \frac{16m^3}{3m + 8} \). 
As $m$ increases, the benefit of using MCGS becomes more significant, as it reduces the number of nodes compared to MCTS.
%

In summary, by sharing nodes and reducing redundant calculations, MCGS significantly reduces the number of nodes and simulations, enabling more efficient search~\cite{DBLP:conf/aips/CzechKK21, shen2018m, saffidine2012ucd}.

\subsection{Optimization Techniques}
\label{sec:optim}

\stitle{Rule-based Pruning.} 
{In the visualization generation process, selecting and expanding nodes are crucial. However, the traditional UCB algorithm may lead to inappropriate node selections due to the lack of visualization-specific knowledge. To alleviate these issues, we introduce a rule-based pruning algorithm. }

{The key aspect of this algorithm is integrating domain knowledge, such as \textit{data transformation rules} and \textit{visualization rules}, to ensure the generated visualizations are syntactically correct.}
For example, after completing a {\sf GROUP BY} operation, if an aggregate function needs to be chosen for the $y$-axis, and the $y$-axis field is categorical, then functions such as {\sf SUM}, {\sf AVG}, and {\sf NONE} become inapplicable. Our algorithm aims to exclude these inappropriate options to enhance the efficiency and accuracy of the search. 

Furthermore, we present a function $L(S, A)$ that takes the current state $S$ and the candidate operation set $A$ as inputs, and outputs a set of operations meeting specific constraints, formalizing the pruning process based on domain-specific knowledge for visualizations.

\stitle{Adaptive Random Exploration Strategy.} {During the visualization generation process, each selected visualization operation (\eg chart type) significantly affects the final visualization. Although the UCB algorithm is designed to make the best choice based on the current state, it may tend to over-explore high-scoring branches as the number of simulations accumulates, overlooking other possibilities. Moreover, the UCB algorithm might favor sub-optimal or inefficient visualization actions, especially when the initial attempts are inaccurately evaluated by the reward function~\cite{DBLP:journals/nature/SilverSSAHGHBLB17, DBLP:conf/aips/CzechKK21}.}

To address these issues, we propose an adaptive random exploration strategy. This strategy adjusts the random selection probability based on the number of already selected clauses, facilitating the transition from exploitation to exploration. This approach increases search space coverage and diversity while avoiding local optima. It mirrors user behavior in constructing visualization queries, with extensive exploration at the initial stage and more targeted selection based on earlier trials at later stages.

The optimized formula for the selection strategy is as follows:

\begin{equation}
a_i = 
\begin{cases} 
random(L(S_i, A_i)), & P_{\text{exploration}} \\
\arg \max_{b \in L(S_i, A_i)} \left[ Q_i(b) + c\sqrt{\frac{2\ln t}{N_i(b)}} \right], & P_{\text{exploitation}}
\end{cases}
\end{equation}
In our proposed decision-making strategy, the final choice $a_i$ for each node $i$ is based on a series of parameters and functions. The set of legal clauses available for node $i$ is represented by $L(S_i, A_i)$. The average reward for each clause $b$ is estimated by $Q_i(b)$, while $N_i(b)$ tracks the number of times clause $b$ has been visited. The total number of visits to the current node is denoted by $t$, and the constant $c$ is used to balance exploration in the process.

Furthermore, the balance between exploration and exploitation in the strategy is controlled by two main probability parameters: $P_{\text{exploration}}$ and $P_{\text{exploitation}}$. The exploration probability, $P_{\text{exploration}}$, is defined as $p_n \alpha^n$, where $p_n$ is the initial probability of making a random choice and $\alpha$ is a constant less than 1 that controls the decay rate of the random selection probability. On the other hand, the exploitation probability, $P_{\text{exploitation}}$, is defined as $1 - p_n \alpha^n$. As the number of clauses $n$ selected in node $i$ increases, the probability of random selection gradually decreases, while the probability of selection based on the UCB algorithm correspondingly increases.

\section{Learning-to-Rate Visualization}
\label{sec:reward}

Unlike environments for games like Atari~\cite{DBLP:journals/nature/SchrittwieserAH20} or Go~\cite{DBLP:journals/nature/SilverHMGSDSAPL16}, which have clear reward and punishment rules, visualization evaluation lacks such clarity and can be biased if based on a single criterion.

To address this, we propose a composite reward function that incorporates visualization best practices, data features, and user preferences to comprehensively evaluate visualization quality.


\subsection{Composite Reward Function}
\label{sec:reward_function}

The composite reward function incorporates \textit{rules of thumb} for visualizations, data features, and user preferences to evaluate the quality of visualization results. The formula for the Composite Reward Function (CRF) is as follows:

\begin{equation}
CRF = S_k \times (\beta S_d + (1 - \beta) S_u)
\end{equation}%
where \( S_k \) represents the assessment result from visualization domain knowledge, \( S_d \) indicates the score based on data features, and \( S_u \) denotes the user preference score. When evaluating a visualization result, we first conduct a preliminary evaluation based on visualization domain knowledge. If the result aligns with the domain knowledge, \( S_k \) is assigned a value of 1; otherwise, it is 0. Clearly, when \( S_k = 0 \) (\ie the visualization result does not align with domain knowledge), the reward value \( CRF \) becomes 0. Conversely, when \( S_k = 1 \), the calculation of \( CRF \) depends on the weighted combination of \( S_d \) and \( S_u \), where the weight coefficient \( \beta \) is introduced to balance the importance between data characteristic scoring and user preference scoring, which is set based on experience.

\subsection{Learn Composite Reward Function}
\label{sec:learn_reward_function}

We leverage well-established, human-annotated visualization corpora~\cite{hu2019vizml, qian2022personalized} and domain knowledge from the visualization community to learn the composite reward function.

\stitle{Leveraging Domain Knowledge.} 
We employ a rule-based method to ensure that the generated visualizations align with best practices and accurately reflect data features. Specifically, this method leverages the data selection, transformation, and visualization rules used by DeepEye~\cite{DBLP:conf/icde/LuoQ0018}.


For example, the rules for pie charts include: (1) Data selection rules: the $x$-axis should represent categorical data, while the $y$-axis should represent numerical data. (2) Data transformation rules: pie charts are unsuitable for data aggregated with the {\small\sf AVG} operation, as they are primarily used to show proportions. (3) Visualization rules: the 
$y$-axis values should not include negatives, and there should be at least two distinct 
$x$-axis values to convey meaningful information. Using this approach, we design 15 rules for different types of visualizations and use these rules as strict constraints to filter out \textit{low-quality} visualizations.


\stitle{Capturing Data Features.} Our model is trained on a dataset annotated by real users~\cite{DBLP:conf/icde/LuoQ0018}, containing 285,236 visualizations and their scores from 42 different domains. For each visualization, we analyze key features, including the data types of the $x$- and $y$-axes, the number of rows, extremes (maximum and minimum values), value diversity (number of different values and the ratio of unique values), and the correlation between the data of the $x$- and $y$-axes and the type of chart. We extract 14 core features for model training.

We use LambdaMART~\cite{wu2008ranking} to train our scoring model, which maps complex relationships between features and scores by building decision trees. Each tree evaluates features, such as the type of data on the $x$-axis, then examines other features, and finally outputs a score. This method enables the model to understand the deep connections between feature combinations and scores. When a new visualization is input, the model computes its features and uses the trained LambdaMART to predict a score.

\stitle{Learning Common User Preferences.} 
{Capturing user preferences for visualization is essential in visualization recommendation systems. Although it is ideal to consider individual preferences in online recommendations, this poses a challenge due to the cold start problem -- new users often lack sufficient interaction history, especially in visualization scenarios. In addition, collecting real-time interaction data and fine-tuning models online for each user can be resource-intensive and time-consuming.

To address these challenges, \textit{our goal is to first learn common user preferences offline} and then make online adjustments to the MCGS algorithm based on feedback from visualization hints, which we will discuss in Section~\ref{sec:recommendation}. This approach strikes a balance between recommending high-quality visualizations and fine-tuning the recommendations to better suit individual users.

To achieve this, our system utilizes the Generative Adversarial Network (GAN)~\cite{goodfellow2020generative}, specifically the IRecGAN~\cite{DBLP:conf/nips/BaiGW19}, to learn common user preferences using a real-world visualization corpus. This corpus includes user interaction logs collected from the Plotly community~\cite{qian2022personalized}. The dataset comprises historical interaction records of 17,469 users, with an average of 5.41 datasets and 1.85 visualizations per user, and each dataset contains an average of 24.39 attributes.
For effective learning of user preferences across datasets and visualizations, we extracted 606 visualization configurations from the original dataset, inspired by~\cite{qian2022personalized}. Each configuration includes dataset-independent design choices such as chart types, colors, and sizes, enabling the model to learn preferences without direct dataset associations and to identify patterns from design choices. By aggregating data, user, and visualization features across different datasets, we generated 15,531 training samples to train the IRecGAN model~\cite{DBLP:conf/nips/BaiGW19}. 

In the first round of visualization recommendations, the evaluations obtained from the trained IRecGAN model serve as a crucial indicator of general user preferences.
In the interaction phase (\eg the second round),  \sys refines and optimizes the recommendation strategy based on the visualization hints selected by users, which will be detailed in Section~\ref{sec:recommendation}.}

\section{Visualization Hints Selection}
\label{sec:recommendation}

While the MCGS-based visualization generation algorithm can recommend high-quality visualizations, it might not always align with individual preferences. Conversely, interactive tools like Voyager2~\cite{wongsuphasawat2017voyager} offer flexible exploration options but can overwhelm users with their complexity. These tools require users to sift through dense control panels and numerous visualization options, much like tackling a challenging ``\textit{fill-in-the-blank}'' puzzle.

\begin{figure}[t!]
	\centering
	\includegraphics[width=1\columnwidth]{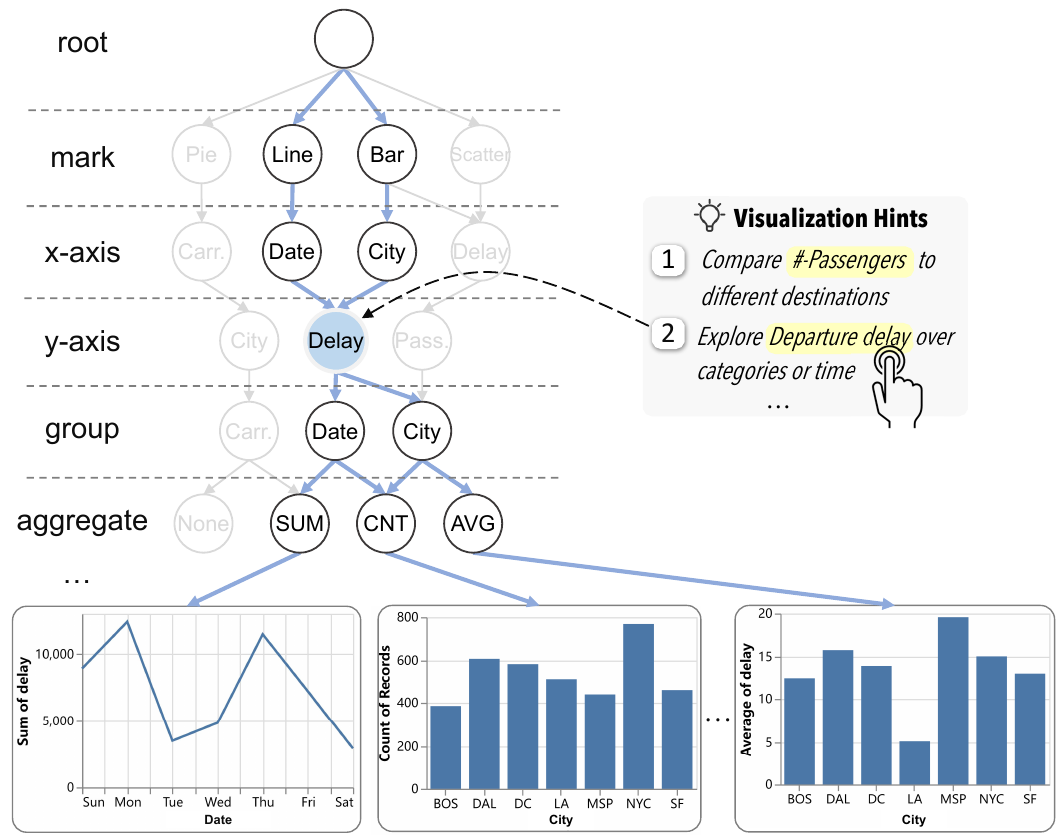}
	\caption{Visualization refinement with visualization hints}
	\label{fig:hint_example}
\end{figure}

To address these issues, we introduce a visualization hints module. Visualization hints represent the high-level visualization intent derived by our system, as shown in Figure~\ref{fig:hint_example}.
This module simplifies the complex decision-making process of visualization creation into straightforward, user-friendly selections, akin to ``\textit{multiple-choice questions}''. This method collects user intent, allowing our MCGS algorithm to be fine-tuned online to better suit individual users.


\begin{definition} [Visualization Hint] A visualization hint \(h\) corresponds to a visualization operation or action, such as selecting data fields, applying aggregate operations, and choosing chart types. It is expressed in easily understandable natural language. Each hint \(h\) is associated with a set of visualizations \(\mathbb{V} = \{v_1, v_2, \ldots, v_n\}\), which are ranked based on the values of the composite reward function to prioritize high-value visualizations. 
\end{definition}

As shown in Figure~\ref{fig:hint_example}, consider a data field-based hint: ``\texttt{Explore $\langle$\textit{data field name}$\rangle$ over categories or time}'',  this hint essentially defines a partial visualization query relevant to the data field.
Therefore, the user can browse these hints and select a desired hint for visualization. Next, \sys will generate a set of relevant visualizations based on the selected hint through the MCGS process.
This approach not only simplifies the steps for the user to define a visualization query but also speeds up the process of gaining data insights by guiding the user through meaningful hints to quickly explore and understand the data.

Our goal is to select the top-$k$ visualization hints that not only cover different aspects and ensure high-quality visualizations but also maintain an appropriate number of visualizations.

\begin{definition} [Top-$k$ Visualization Hints Selection]
Given a set of hints \( \mathbb{H} = \{h_1, h_2, \ldots, h_n\} \), where each hint \( h_i \) is associated with a set of visualizations \( \mathbb{V}_i \), and each visualization \( v \in \mathbb{V}_i \) has an associated reward value \( r_v \), the goal is to select a subset \( \mathbb{H'} \subseteq \mathbb{H} \) consisting of \( k \) hints. The selection must maximize the total reward value of the subset \( \mathbb{H'} \), while ensuring the total count of visualizations in \( \mathbb{H'} \) does not exceed a predefined budget \( B \). The optimization problem can be formulated as follows:
\begin{equation}
    \text{Maximize} \quad F(\mathbb{H'}) = \sum_{h_i \in \mathbb{H'}} \sum_{v \in \mathbb{V}_i} r_v
\end{equation}
\begin{equation}
    \text{Subject to} \quad \sum_{h_i \in \mathbb{H'}} |\mathbb{V}_i| \leq B \quad \text{and} \quad |\mathbb{H'}| = k
\end{equation}%
where \( |\mathbb{V}_i| \) represents the number of visualizations associated with hint \( h_i \), \( F(\mathbb{H'}) \) is the total reward value of the selected subset \( \mathbb{H'} \), \( |\mathbb{H'}| \) is the number of selected hints, and \( B \) is the upper limit on the budget for the number of visualizations.
\end{definition}

However, the problem of selecting hints is  NP-hard because it is equivalent to a known NP-hard problem -- the Budgeted Maximum Coverage problem~\cite{DBLP:journals/ipl/KhullerMN99}. To address this, we propose an efficient algorithm to select the top-$k$ visualization hints. 
To achieve effective hint selection, the first step is to construct a comprehensive set of hints and evaluate the benefit of each hint. The purpose of this initial phase is to ensure that in the subsequent selection process, the best hint can be selected from this evaluated set, thereby maximizing the reward value. In this way, the algorithm mainly consists of two steps: candidate hints generation and top-$k$ hints selection.

\begin{figure}[t!]
	\removelatexerror
	{\small
	\begin{algorithm}[H]
		\caption{Top-$k$ Visualization Hints Selection \label{alg:topk_hint_selection}}
		\KwIn{Set of hints $\mathbb{H} = \{h_1, h_2, \ldots, h_n\}$, $B$, $k$;}
		\KwOut{Selected top-k set of hints $\mathbb{H}'$;}
		$\mathbb{H}' \leftarrow \emptyset$; $totalCost \leftarrow 0$; \\
		$\mathbb{H}_v \leftarrow \{h \in \mathbb{H} \mid |h| \leq B\}$; // 1. Filter valid hints \\
		
		$\mathbb{H}_v \leftarrow \text{SortByScore} (\mathbb{H}_v)$; // 2. Sort hints by score \\
		// 3. Selection of top-k hints \\
		\For{each $h_i$ in $\mathbb{H}_v$}{
			\If{$|\mathbb{H}'| < k$ and $totalCost + |h_i| \leq B$}{
				$\mathbb{H}'.\text{append} (h_i)$;\\
				$totalCost \leftarrow totalCost + |h_i|$;
			}
			\If{$|\mathbb{H}'| = k$}{
				break;
			}
		}
		\Return $\mathbb{H}'$;
	\end{algorithm}
}
\end{figure}

\stitle{Candidate Visualization Hints Generation.} We identify high-value nodes (\ie visualization operations) from our visualization query graph to pick the candidate hints. 
When computing the reward for each hint, we introduce a decay coefficient $\delta$ to adjust the scoring weight of each visualization, considering that the same visualization appearing in different hints may reduce its uniqueness. The decay coefficient $\delta$, is defined as:
    $\delta = \log\left(\frac{N_{\text{total}}}{N_{\text{viz}}}\right)$,
where $N_{\text{total}}$ is the total number of hints and $N_{\text{viz}}$ is the number of hints containing the specific visualization. This coefficient reflects the frequency of visualizations, with higher decay coefficients for more frequent visualizations, thereby reducing their overall scores. After multiplying this score by the reward value of the visualization, its final value is determined. Therefore, after collecting all the hints, we calculate the frequency of the appearance of each visualization and combine it with the decay coefficients to compute its score. 

\stitle{Top-$k$ Visualization Hints Selection.} 
Our goal is to select the best subset of hints that maximizes the total reward within a predefined budget $B$. Algorithm~\ref{alg:topk_hint_selection} shows the pseudo-code.
It first selects all hints with a cost not exceeding the budget $B$, forming a candidate hint set (Line 2). Then, it sorts valid hints based on the corresponding average visualization score computed by the composite reward function (Line 3). The algorithm continues to go through this sorted set, picking hints to add to the final selected set until one of two conditions is met: the number of selected hints reaches the predefined $k$, or adding more hints would cause the total cost to exceed the budget $B$ (Line 5-10). In this way, the algorithm prioritizes high-scoring hints while maintaining the budget constraint, thus maximizing the total reward value under the given budget.

\stitle{User Feedback for Refinement.} \sys leverages user-selected visualization hints to guide the node exploration strategy during the MCGS process. When users select a specific hint, they direct the search graph to expand in a certain direction.  For example, as shown in Figure~\ref{fig:hint_example}, if a user clicks on the hint ``{\sf Explore {\tt delay} over categories or time}'', the system focuses on applying the data field ``{\tt delay}'' to the $y$-axis. During the search process, the system freezes other nodes and explores only those related to ``{\tt delay}'', ensuring the generated results are relevant to the target field. \textit{Thus, this method can align the search process with user preferences and effectively prune the search space, enhancing the efficiency and accuracy of MCGS-based visualization recommendations.}

\section{Experiments}
\label{sec:experiment}

\begin{table}[t!]
	\caption{{Statistics of the experimental datasets (Vis.: Charts)}}
	\label{tab:experimental_datasets}
    \scalebox{0.7}{
        \begin{tabular}{c|c|c|c|c|c|c}
            \hline
            Datasets & {\#-Tables} & \#-Vis. & Avg(\#-Vis.) & Avg(\#-Rows) & Avg(\#-Col.) & Max(\#-Col.)  \\ \hline \hline
            VizML & {79,475} & 162,905 & 2 & 2,817.8 & 3.3 & 25  \\ \hline
            KaggleBench & {8} & 252 &  31.5 & 32,585.9 & 9.1 & 15  \\ \hline
        \end{tabular}
    }
\end{table}

\begin{figure}[t] 
	\centering \includegraphics[width=1.0\columnwidth]{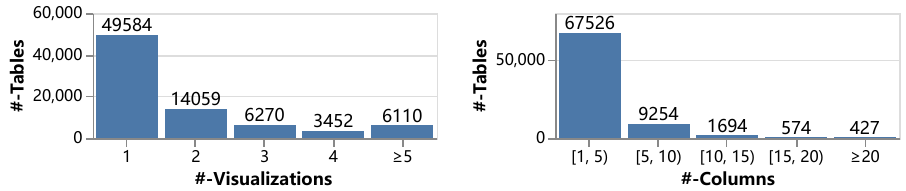} 
	\vspace{-1.5em}
	\caption{{Statistics of the VizML dataset}} 
	\label{fig:vizml_statistical} 
\end{figure}

\begin{table}[t!]
	\centering
	\caption{Details in KaggleBench dataset}
	\label{tab:multi-turn_datasets}
	\scalebox{0.9}{
		\begin{tabular}{c||c|c|c|c}
			\hline
			& Datasets & \#-Rows & \#-Columns & \#-Vis. \\ \hline \hline
			D1 & StudentPerformance & 1,000 & 8 & 34 \\ \hline
			D2 & AirplaneCrashes & 5,191 & 6 & 22 \\ \hline
			D3 & VideoGame & 6,825 & 15 & 48 \\ \hline
			D4 & GooglePlayStore & 9,659 & 11 & 33 \\ \hline
			D5 & AvocadosPrice & 18,249 & 7 & 20 \\ \hline
			D6 & SuicideRates & 27,820 & 9 & 25 \\ \hline
			D7 & Zomato & 51,717 & 11 & 49 \\ \hline
			D8 & GunViolence & 140,226 & 6 & 21 \\ \hline
		\end{tabular}
	}
\end{table}

\begin{table*}[t!]
	\centering
	\caption{{Effectiveness of the first round of visualization recommendation}}
	\label{tab:init_rec_results}
	\scalebox{0.8}{
		\begin{tabular}{c|c|c|c|c|c|c|c|c|c|c|c}
			\hline
			\multirow{2}{*}{D} & \multirow{2}{*}{{Tasks}} & \multirow{2}{*}{Metrics} & \multicolumn{6}{c|}{The State-of-the-Art Methods} & \multicolumn{3}{c}{Our Methods} \\ \cline{4-12}
			&  &  & Data2Vis~\cite{dibia2019data2vis} & VizGRank~\cite{DBLP:conf/dasfaa/GaoHJZW21} & DeepEye~\cite{DBLP:conf/icde/LuoQ0018} & PVisRec~\cite{qian2022personalized} & VizML~\cite{hu2019vizml} & {LLM4Vis~\cite{DBLP:conf/emnlp/WangZWLW23}} & MCTS & {\sys-} & \sys \\ \hline \hline
			\multirow{6}{*}{\rotatebox[origin=c]{90}{VizML}} & \multirow{2}{*}{Data Queries} & Hit@1 & 47.5\% & 57.6\% & 52.4\% & 52.3\% & - & - & 78.3\% & {\textbf{79.7\%}} & 79.3\% \\ \cline{3-12}
			&  & Hit@3 & 51.3\% & 67.2\% & 67.6\% & 58.7\% & - & - & 88.2\% & {91.3\%} & \textbf{91.9\%} \\ \cline{2-12}
			& \multirow{2}{*}{Design Choices} & Hit@1 & 41.7\% & 34.9\% & 34.1\% & 28.9\% & 28.7\% & {47.9\%} & 42.4\% & {\textbf{50.6\%}} & 48.7\% \\ \cline{3-12}
			&  & Hit@3 & 43.7\% & 42.9\% & 40.7\% & 51.3\% & - & - & 77.1\% & {\textbf{81.8\%}} & 81.5\% \\ \cline{2-12}
			& \multirow{2}{*}{Overall} & Hit@1 & 24.3\% & 25.6\% & 25.7\% & 21.8\% & - & - & 33.1\% & {\textbf{37.9\%}} & 36.9\% \\ \cline{3-12}
			&  & Hit@3 & 26.9\% & 30.1\% & 33.9\% & 42.3\% & - & - & 64.7\% & {\textbf{68.4\%}} & 67.4\% \\ \hline \hline
			\multirow{6}{*}{\rotatebox[origin=c]{90}{KaggleBench}} & \multirow{2}{*}{Data Queries} & P@10 & 41.2\% & 58.7\% & 62.5\% & 42.5\% & - & - & 52.2\% & {60.0\%} & \textbf{63.8\%} \\ \cline{3-12}
			&  & R10@30 & 25.0\% & 50.0\% & 48.7\% & 67.5\% & - & - & 73.6\% & {80.1\%} & \textbf{83.7\%} \\ \cline{2-12}
			& \multirow{2}{*}{Design Choices} & P@10 & 88.7\% & 87.5\% & 93.7\% & 91.9\% & Hit@2:78.3\% & {Hit@2:87.6\%} & 93.8\% & \textbf{{96.3\%}} & \textbf{96.3\%} \\ \cline{3-12}
			&  & R10@30 & 95.0\% & 81.3\% & 95.0\% & 85.0\% & - & - & 92.5\% & \textbf{{96.2\%}} & \textbf{96.2\%} \\ \cline{2-12}
			& \multirow{2}{*}{Overall} & P@10 & 28.7\% & 43.7\% & 48.7\% & 36.7\% & - & - & 45.4\% & {51.3\%} & \textbf{55.0\%} \\ \cline{3-12}
			&  & R10@30 & 13.8\% & 41.3\% & 33.7\% & 60.0\% & - & - & 63.8\% & {72.5\%} & \textbf{74.9\%} \\ \hline
		\end{tabular}
	}
\end{table*}

\subsection{Experiment Settings}
\label{sec:exp_settings}

\stitle{Datasets.}  Table~\ref{tab:experimental_datasets} shows 
two real-world datasets for experiments.

\stab (1) {\sf VizML~\cite{hu2019vizml}}, derived from the Plotly community, features around 120,000  dataset-visualization pairs created by real users. This dataset was refined by removing entries missing table or chart data and cleaning up invalid characters. The dataset has four types of charts, namely bar, pie, line, and scatter visualizations, resulting in 79,475 valid dataset-visualization pairs. These were randomly divided into training, validation, and testing sets in a 7:1:2 ratio, allocating 55,632 pairs for training, 7,947 for validation, and 15,896 for testing. 
{The statistical information of the dataset, shown in Figure~\ref{fig:vizml_statistical}, indicates a broad coverage, including various tables and visualizations used by users across different domains and tasks.} 

\stab (2)  {\sf KaggleBench~\cite{DBLP:conf/dasfaa/GaoHJZW21}} is a public benchmark designed to evaluate the effectiveness of visualization recommendations. Its data mainly come from numerous data competitions and the corresponding visualizations provided by the Kaggle platform. This dataset was refined by filtering out low-quality datasets, removing rows with missing data, and fixing invalid characters. Finally, we obtained 8 datasets for evaluation, as shown in Table~\ref{tab:multi-turn_datasets}.

We trained or configured all methods using the VizML dataset and tested them on the two datasets mentioned above.

\stitle{Methods.} We evaluate the following methods.

\etitle{(I) AI-powered Visualization Methods}:

 (1) {\sf Data2Vis~\cite{dibia2019data2vis}} transforms datasets into visualization queries using a sequence-to-sequence model.

 (2) {\sf VizGRank~\cite{DBLP:conf/dasfaa/GaoHJZW21}}  achieves visualization recommendation by modeling the relationships between visualizations as graphs and using graph-based ranking algorithms.

 (3) {\sf DeepEye~\cite{DBLP:conf/icde/LuoQ0018}} combines data features with domain knowledge to recommend top-$k$ good visualizations.

 (4) {\sf PVisRec~\cite{qian2022personalized}} recommends a set of visualizations by learning from user preferences. 

 (5) {\sf VizML~\cite{hu2019vizml}} uses deep learning models for visualization recommendations. It focuses on five specific types of tasks, excluding data querying. Our evaluation will assess VizML's effectiveness in recommending design choices.

{ (6)  
	{\sf LLM4Vis~\cite{DBLP:conf/emnlp/WangZWLW23}}
	 uses in-context learning to interact with ChatGPT for recommending visualizations. Like VizML, it excludes data querying tasks. Thus, our comparison focuses on design choices, evaluated using the same Hit@$k$ metric as LLM4Vis.}

\etitle{(II) Human-powered Visualization Methods:}

 (7) {\sf Voyager2~\cite{wongsuphasawat2017voyager}} is an interactive system that allows users to create and explore visualizations through click-based interactions.

\etitle{(III) Human and AI paired Visualization Methods:}

 (8) {\sf \sys} (ours) is the full implementation based on MCGS and the composite reward function, as described in this paper.

{ (9)  
	{\sf \sys-} (ours) differs from \sys in the composite reward function, where the capturing data features model is trained on the VizML corpus.}

{ (10) {\sf LLM4Vis+ (ours)}
is an improved version of LLM4Vis~\cite{DBLP:conf/emnlp/WangZWLW23}. We enhanced LLM4Vis to support data queries and top-$k$ visualizations.}

 (11) {\sf MCTS-based Baseline} (ours) follows the same implementation as \sys, except it uses the MCTS-based method for recommending visualizations.


\stitle{Metrics.} 
Following previous studies evaluating automatic visualization systems~\cite{zhou2021table2charts, dibia2019data2vis, hu2019vizml, DBLP:journals/pacmmod/LuoZ00CS23}, we employ Hit@$k$, P@$k$, and Rt@$k$ as evaluation metrics. Given that creating a visualization involves data queries, design choices, and final integration, our evaluation is divided into three tasks. The metrics are defined as follows:

\stab (1) \textit{Hit@$k$:} This metric evaluates whether the ground truth appears in the top-$k$ results. We apply this to the VizML dataset with $k$ set to 3, as each user typically creates about 2 visualizations on average.

\stab (2) \textit{P@$k$:} This metric measures how many ground truths are in the top-$k$ results. For the KaggleBench dataset, we set $k$ to 10, considering that each dataset typically contains over 20 visualizations.

\stab (3) \textit{R$t$@$k$:} This metric assesses how many of the top-$t$ ground truths are covered in the top-$k$ results. We use R10@30 for the KaggleBench dataset, analyzing how many of the top 10 ground truths are included in the first 30 results returned by the model.

\stitle{Experimental Environment.} Experiments were performed on an Ubuntu 22.04 Server LTS with dual Intel Xeon 8383C CPUs, 512GB RAM, and eight NVIDIA RTX 4090 GPUs.

\begin{figure*}[t!]
	\begin{minipage}{\textwidth}
		\centering
		\begin{minipage}{0.62\textwidth}
			\begin{table}[H]
				\centering
				\caption{{Performance \vs \#-Iterations on KaggleBench}}
				\label{tab:multi-turn_results}
				\vspace{-1em}
				\scalebox{0.65}{
					\begin{tabular}{c||c|c|c|c|c|c|c|c|c|c|c}
						\hline
						\multirow{2}{*}{Tasks} & \multirow{2}{*}{Metrics} & \multicolumn{1}{c|}{DeepEye}  & \multicolumn{3}{c|}{Voyager2} & \multicolumn{3}{c|}{LLM4Vis+ (ours)} & \multicolumn{3}{c}{\sys (ours)} \\ \cline{3-12} 
						&  & Iter. 1  & Iter. 1  & Iter. 2 & Iter. 3 & Iter. 1  & Iter. 2 & Iter. 3 & Iter. 1  & Iter. 2 & Iter. 3 \\ \hline \hline
						Data Queries & P@10 & 62.5\%  & 45.0\% & 55.1\% & 58.0\% & 47.2\% & 65.0\% & 74.3\% & 63.8\% & 69.5\% & \textbf{79.2\%} \\ \cline{1-12}
						Design Choices & P@10 & 93.7\% & 78.7\% & 96.3\% & 97.4\% & 75.6\% & 91.5\% & 97.2\% & 96.3\% & 97.6\% & \textbf{99.3\%} \\ \cline{1-12}
						Overall & P@10 & 48.7\% & 40.0\% & 44.9\% & 45.7\% & 41.0\% & 55.6\% & 65.3\% & 55.0\% & 58.2\% & \textbf{68.8\%} \\ \hline
					\end{tabular}
				}
			\end{table}

		\end{minipage}
		\begin{minipage}{0.37\textwidth}
			\begin{table}[H]
				\centering
				\caption{{Effectiveness of Hints Selection}}
				\label{tab:hit_select_results}
				\vspace{-1em}
				\scalebox{0.8}{\begin{tabular}{c||c|c|c|c}
						\hline
						Dataset & Metrics & Round 1 & Round 2 & Round 3 \\ \hline \hline
						\multirow{3}{*}{KaggleBench} & Hit@1 & 64.7\% & 65.2\% & \textbf{69.1\%} \\ \cline{2-5}
						& Hit@3 & 79.4\% & 82.2\% & \textbf{85.7\%} \\ \cline{2-5}
						& Hit@5 & 88.2\% & 89.6\% & \textbf{92.1\%} \\ \hline
				\end{tabular}}
			\end{table}
		\end{minipage}
	\end{minipage}
\end{figure*}

\begin{figure*}[t!]
	\centering
	\begin{minipage}{0.585\textwidth}
		\centering 
		\includegraphics[width=1.02\textwidth]{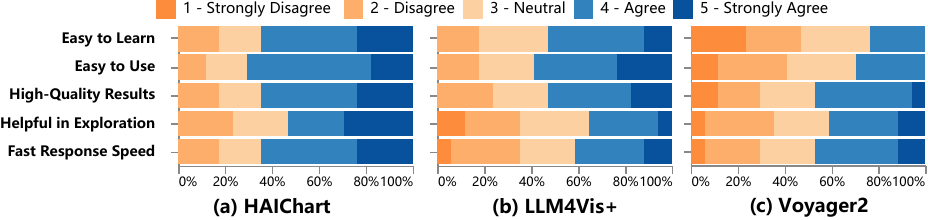}
		\vspace{-2em}
		\caption{{Qualitative analysis on user study}} 
		\label{fig:user_study} 
	\end{minipage}
	\begin{minipage}{0.41\textwidth}
		\centering 
		\includegraphics[width=0.95\textwidth]{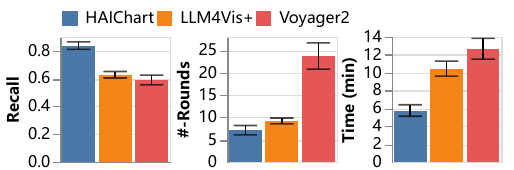}
		\caption{{Quantitative analysis on user study}} 
		\label{fig:user_inter} 
	\end{minipage}
	\vspace{-.5em}
\end{figure*}

\subsection{Experimental Results}
\subsubsection{Exp-1: Effectiveness of the First-round of Recommendations.}
This experiment evaluates the effectiveness of the first-round visualization recommendations by \sys. We tested all methods on the VizML and KaggleBench datasets. 
Table~\ref{tab:init_rec_results} shows the results.

\stab (1) {Overall, our methods (\sys and \sys-) significantly outperform all state-of-the-art methods across all metrics, showing the effectiveness of our framework.
The performance gap between \sys- and \sys is primarily due to the capturing data features module: \sys- is trained on the VizML training set, while \sys is trained on a more diversified corpus~\cite{DBLP:conf/icde/LuoQ0018}.
}

\stab (2) On the VizML dataset, \sys achieves 36.9\% Hit@1 and 67.4\% Hit@3 on the overall task, surpassing the competitive methods DeepEye and PVisRec by 11.2\% and 25.1\%, respectively. On the KaggleBench dataset, \sys achieves 55\% P@10 and 74.9\% R10@30 on the overall task, outperforming DeepEye and PVisRec by 6.3\% and 14.9\%. These results demonstrate \sys's effectiveness in leveraging MCGS algorithms with a composite reward function to find high-quality visualizations in a large search space.

\stab (3) DeepEye performs well on the P@10 metric using a rule-based approach, while PVisRec excels on the Rt@30 metric with personalized recommendations. By blending visualization rules, user preferences, and data features, \sys achieves strong results in both the P@10 and Rt@30 metrics.

\subsubsection{Exp-2: Effectiveness of Multi-round Recommendations.}
\label{sec:exp_multi_round} 
This experiment assesses whether \sys improves visualization recommendations and enhances user efficiency in data exploration through multiple rounds of interaction and visualization hints.

\etitle{\sf Participants.} 
The study involved 17 participants (7 females, 10 males, aged 21-33), including 12 experts (6 Ph.D. candidates, 4 master's students, and 2 undergraduates in computer science) and 5 non-experts from non-technical backgrounds. 

\etitle{\sf Task.}
Participants used \sys, DeepEye, Voyager2, and LLM4Vis+ to analyze eight datasets from KaggleBench. These datasets required specific analytical tasks, such as analyzing the relationship between student performance and background factors in the StudentsPerformance dataset, as well as open-ended explorations to evaluate each tool's adaptability and efficiency.

\etitle{\sf Procedure.} 
{\sf (1)  Preparation}: Participants were introduced to the datasets and tools through a hands-on demonstration.
%
{\sf (2) Experimentation}: 
Participants explored datasets using these tools, during which we recorded their time spent, interactions, and visualizations. After each round, \sys offered the top-$9$ hints and new recommendations.
%
{\sf (3)  User Feedback}: 
After the experiment, participants rated each tool on a five-point Likert scale~\cite{allen2007likert, li2024coinsight} and provided feedback through interviews.

\etitle{\sf User Study Results.} {We conducted both quantitative and qualitative analyses. The observations are as follows:}

\stab (1) 
As shown in Table~\ref{tab:multi-turn_results}, \sys improves visualization recommendations through user feedback. After two rounds of interactions, P@10 increased from 55\% to 68.8\%, outperforming LLM4Vis+ and Voyager2 by 3.5\% and 23.1\%, respectively. Unlike Voyager2's manual exploration and LLM4Vis+'s time-consuming natural language queries, \sys's hints guide users efficiently. DeepEye, achieving 48.7\% in the first round, lacks multi-turn recommendation support and does not improve with additional interactions.

\stab (2) 
According to Figure~\ref{fig:user_inter}, \sys is 1.8 and 2.2 times faster than LLM4Vis+ and Voyager2, respectively, achieving an average recall of 83.7\%. This is 21\% and 24.8\% higher than LLM4Vis+ and Voyager2, indicating that \sys meets analytical needs with fewer interactions and less time. Furthermore, user feedback (Figure~\ref{fig:user_study}) underscores \sys's ease of learning and usability, facilitating more effective exploration and higher quality visualizations.

\stab (3) 
Feedback from participants highlighted \sys's practicality in data exploration. They widely acknowledged that \sys simplifies tasks and enhances efficiency with helpful hints. For example, Participant P1 stated, ``\textit{Unlike Voyager2, which is not user-friendly for beginners, \sys makes it easier to choose data fields and operations through intuitive hints.}'' 
Participant P5 mentioned a limitation with LLM4Vis+, ``\textit{Natural language interaction simplifies creating initial visualizations, but fine-tuning is challenging. LLMs often misunderstand my intent, requiring repeated adjustments.}'' This suggests that natural language ambiguity may cause `\textit{communication}' issues with LLMs during data exploration. Intuitive control panels or guidance, like hints, would enhance the user experience.

Nevertheless, participants also suggested improvements for \sys, such as adding a control panel for enhanced flexibility (Participant P10) and supporting more complex chart types to broaden analysis capabilities (Participant P15).

\subsubsection{Exp-3: Effectiveness of Visualization Hints Selection.}
This experiment evaluates the effectiveness of visualization hint selection by analyzing interaction logs from 17 users in a multi-round recommendation experiment (Exp-2), using the Hit@$k$ metric.

The results in Table~\ref{tab:hit_select_results} show that Hit@1 is 64.7\% in the first round of recommendations, demonstrating the system's ability to accurately predict the hint most interesting to the user. As the number of interaction rounds increases, the recommendation accuracy improves further. By the third round, the Hit@3 value increased to 85.7\%. These results demonstrate the system's effectiveness in prioritizing the information required by the user and the effectiveness of the hint selection in guiding the user's decision-making process.

\begin{table}[t!]
	\centering
	\caption{{Ablation studies on \sys (overall performance)}}
	\label{tab:ablation_study}
	\scalebox{0.72}{
	\begin{tabular}{c|c||c|c|c|c}
		\hline
		\multicolumn{2}{c||}{\multirow{2}{*}{Methods}} & 
		\multicolumn{2}{c|}{VizML} & \multicolumn{2}{c}{KaggleBench} \\ 
		\cline{3-6} 
		\multicolumn{2}{c||}{} & Hit@1 & Hit@3 & P@10 & R10@30 \\ \hline \hline
		\multicolumn{2}{c||}{\sys} & \textbf{36.9\%} & \textbf{67.4\%} & 
		\textbf{55.0\%} & \textbf{74.9\%} \\ \hline \hline
		
		\multirow{2}{*}{Opt. Tech.} & w/o Rule-based 
		Pruning 
		& 34.6\% & 65.3\% & 
		37.4\% & 40.0\% \\ \cline{2-6} 
		& w/o Adapt. Random Exploration & 33.4\% & 42.1\% & 45.7\% & 65.0\% 
		\\ \hline \hline
		\multirow{3}{*}{\shortstack{Composite \\ Reward Func.}} & w/o Domain 
		Knowl. & 26.3\% 
		& 60.2\% & 31.0\% & 
		33.8\% \\ \cline{2-6} 
		& w/o User Preferences & 30.8\% & 64.2\% & 40.7\% & 54.9\% \\ 
		\cline{2-6} 
		& w/o Data Features & 34.2\% & 64.1\% & 36.0\% & 68.7\% \\ \hline
	\end{tabular}}
\end{table}

\subsubsection{{Exp-4: Ablation Study of the MCGS Optimization Techniques}} 

{We proposed several MCGS optimization techniques, including Rule-based Pruning and an Adaptive Random Exploration Strategy. We conducted ablation studies using the VizML and KaggleBench datasets to evaluate their impact on overall performance.}

{The results in Table~\ref{tab:ablation_study} show that removing any optimization technique from either dataset leads to a decline in performance, confirming their importance. Specifically, in the VizML dataset, removing the Adaptive Random Exploration Strategy reduces the improvement from Hit@1 to Hit@3 from 30.5\% to 8.7\%, which demonstrates that this strategy enhances result diversity and avoids local optima, thereby improving performance.}

\subsubsection{Exp-5: Ablation Study of the Composite Reward Function}

Our composite reward function consists of three components: data features, visualization domain knowledge, and user preferences. 
%
We conducted ablation studies using the VizML and KaggleBench datasets to evaluate their impact on overall performance.

Table~\ref{tab:ablation_study} shows the results with the following observations.

Overall, removing any component decreased the performance of \sys.
Performance drops were more significant when removing domain knowledge or user preferences, indicating that these aspects are crucial for aligning visualizations with user needs. In summary, integrating data features, visualization domain knowledge, and user preferences in the composite reward function enhances the effectiveness of visualization recommendations.

\begin{figure}[t!]
	\centering
\includegraphics[width=1\columnwidth]{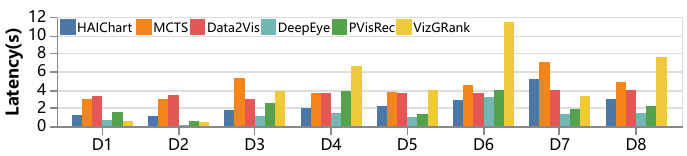}
\vspace{-2em}
	\caption{{Efficiency on KaggleBench datasets}}
	\label{fig:efficiency_results}
 \vspace{-1em}
\end{figure}

\subsubsection{Exp-6: The Efficiency of \sys.}
We compared the efficiency of \sys and existing methods using the KaggleBench dataset, which includes eight tables (D1 to D8) of varying sizes.

Figure~\ref{fig:efficiency_results} shows the results with the following observations:

\stab(1) \sys generates visualizations in 1 to 5 seconds, with an average of 2.4 seconds, and hint generation takes only 1 to 3 milliseconds, validating its capability for rapid visualization exploration.

\stab(2) 
{\sys is, on average, 1.8 times faster than MCTS, especially for datasets with many columns. For example, with the D3 dataset (15 columns), \sys's graph-based structure outperforms MCTS's tree-based structure by 2.9 times.}

%
Overall, \sys is efficient enough for data visualization.




\section{Related Work}
\label{sec:related}

\stitle{Human-powered Visualization} tools such as Tableau, Voyager, and Polaris~\cite{tableau, excel, wongsuphasawat2015voyager, stolte2008polaris, DBLP:conf/cidr/WuPMZR17, DBLP:journals/cgf/SatyanarayanH14} allow users to select or adjust data sources, chart types, and data transformation operations for data visualization. For example, Tableau~\cite{tableau} offers a code-free interface for creating visualizations through click-and-drag actions. However, these tools are highly dependent on user skills, presenting challenges such as steep learning curves.

\stitle{AI-powered Automatic Visualization} tools~\cite{DBLP:conf/icde/LuoQ0018, hu2019vizml, li2021kg4vis, zhou2021table2charts, DBLP:conf/dasfaa/GaoHJZW21, qian2022personalized, DBLP:journals/tkde/LuoQCTLL22, DBLP:conf/sigmod/LuoQ00W18, DBLP:conf/icde/LuoCQ0020, qin2018deepeye, DBLP:conf/sigmod/Chai0FL23} uses algorithms to automatically generate and recommend meaningful visualizations. 
For example, DeepEye~\cite{DBLP:conf/icde/LuoQ0018} uses machine learning to recommend good visualizations based on data features and domain knowledge. 
Similarly, VizML~\cite{hu2019vizml} trains a neural network model to recommend charts based on large-scale real-world visualization cases.
PVisRec~\cite{qian2022personalized} suggests personalized visualizations based on the user's past interactions. 
%
However, these systems may overlook user intent, resulting in recommended visualizations that may not always meet users' needs.

\stitle{Large Language Models for Visualization (LLM4VIS).} LLM4VIS leverages large language models to transform natural language queries into data visualizations~\cite{ncnet, DBLP:conf/sigmod/Luo00CLQ21, DBLP:conf/vis/Luo00CLQ21, DBLP:journals/corr/abs-2112-12926,  wang2022interactive, tian2023chartgpt, DBLP:conf/emnlp/WangZWLW23, DBLP:conf/sigmod/TangLOLC22}.
For example, ChartGPT~\cite{tian2023chartgpt} employs LLMs to generate visualizations from natural language queries, while LLM4Vis~\cite{DBLP:conf/emnlp/WangZWLW23} uses few-shot learning to suggest visualization types and explain them in text. These methods rely on users providing clear query descriptions and may struggle with accurately modifying visualizations through natural language~\cite{DBLP:journals/corr/abs-2406-07815}.
Our method complements LLM4VIS by providing visualization hints and easy-to-adjust features, reducing user input and making the visualization creation process more efficient.

\stitle{Reinforcement Learning} is a learning approach based on trial-and-error, where an agent receives feedback from its environment~\cite{DBLP:conf/icde/LiuCLLFT22}.
It has been successfully used in various tasks such as item recommendation~\cite{DBLP:conf/aaai/WangZDTWFCC21} and data preparation~\cite{DBLP:conf/sigmod/Chai0FL23, DBLP:journals/pvldb/ChaiLTLL22, DBLP:journals/tkde/ChaiWLNL23}.
Recently, it has also been applied to visualization tasks~\cite{zhou2021table2charts, DBLP:conf/sigmod/Chen022, wu2020mobilevisfixer, deng2022dashbot}. 
For example, PI2~\cite{DBLP:conf/sigmod/Chen022, DBLP:conf/sigmod/TaoC022} 
uses MCTS to generate interactive UI widgets from SQL logs, assisting developers in understanding necessary queries for analysis tasks. Unlike PI2, which requires user-provided SQL queries, \sys automatically recommends visualizations for datasets and offers hints to guide data exploration. 
Both systems facilitate data exploration but differ in their application scenarios.



\section{Conclusion}
\label{sec:conclusion}
We introduce \sys, which pairs human insight with AI capabilities to enhance visualization quality \textit{progressively} and \textit{iteratively} through user feedback.
\sys utilizes a Monte Carlo Graph Search-based visualization algorithm for automatically recommending high-quality visualizations.
It is also equipped with a visualization hints mechanism to actively incorporate the user feedback and thus fine-tune the visualization generation algorithm iteratively. 
\sys has been validated for its effectiveness in creating high-quality visualizations that align with user preferences. {Future research could explore integrating LLMs into our framework to generate and evaluate visualizations, potentially enhancing generalizability and robustness for diverse applications.}


\begin{acks}
This paper was supported by Guangdong Basic and Applied Basic Research Foundation (2023A1515110545), National Key R\&D Program of China (2023YFB4503600), NSF of China (61925205, 62232009, 62102215),  Zhongguancun Lab, Huawei, TAL education, Beijing National Research Center for Information Science and Technology (BNRist), CCF-Huawei Populus Grove Fund (CCF-HuaweiDB202306), and the Fundamental Research Funds for the Central Universities.
\end{acks}

\clearpage
\balance
\bibliographystyle{ACM-Reference-Format}
\bibliography{main}
\end{document}